\documentclass[12pt,preprint]{aastex}

\newcommand{\chicold}{\chi_\mathrm{cold}}
\newcommand{\gchicold}{\left(g\chi\right)_\mathrm{cold}}
\newcommand{\gchinought}{\left(g\chi\right)_0}
\newcommand{\Gammabar}{\bar{\Gamma}}
\newcommand{\Pbar}{\bar{P}}
\newcommand{\Nbar}{\bar{N}}
\newcommand{\gbar}{\bar{g}}
\newcommand{\fbar}{\bar{f}}
\renewcommand{\hbar}{\bar{h}}

\begin{document}

\title{Composite self-similar solutions for relativistic shocks:\\ the transition to cold fluid temperatures}

\author{Margaret Pan$^{1,2}$ and Re'em Sari$^{2,3}$}

\altaffiltext{1}{School of Natural Sciences, Institute for Advanced Study, Princeton, NJ 08540}
\altaffiltext{2}{130-33 Caltech, Pasadena, CA 91125}
\altaffiltext{3}{Racah Institute of Physics, Hebrew University, Jerusalem 91904, Israel}

\begin{abstract}
The flow resulting from a strong ultrarelativistic shock moving
through a stellar envelope with a polytrope-like density profile has
been studied analytically and numerically at early times while the
fluid temperature is relativistic---that is, just before and just
after the shock breaks out of the star. Such a flow should expand and
accelerate as its internal energy is converted to bulk kinetic energy;
at late enough times, the assumption of relativistic temperatures
becomes invalid.  Here we present a new self-similar solution for the
post-breakout flow when the accelerating fluid has bulk kinetic
Lorentz factors much larger than unity but is cooling through $p/n$ of
order unity to subrelativistic temperatures. This solution gives a
relation between a fluid element's terminal Lorentz factor and that
element's Lorentz factor just after it is shocked. Our numerical
integrations agree well with the solution. While our solution assumes
a planar flow, we show that corrections due to spherical geometry are
important only for extremely fast ejecta originating in a region very
close to the stellar surface. This region grows if the shock becomes
relativistic deeper in the star.
\end{abstract}

\section{Introduction} \label{introduction}

The energy and Lorentz factor that we expect in the ejecta in
supernovae and gamma-ray bursts are important because they constrain
the amount of energy that can be deposited in the photons we observe
from these explosions. Previous work on the ejecta, notably
\cite{tan01}, uses as a starting point the analytic solutions of
\cite{johnson71} for a planar relativistic shock propagating into cold
surroundings: by the time the shock reaches the outer envelope of the
star, the likely source of the ejecta, it has accelerated to
relativistic speeds and its geometry is planar. Several authors have
described the reltivistic shock's propagation and acceleration through
this envelope \citep[see, for example,][]{best00,perna02,sari05}, but
the work of \cite{johnson71} and other analytic work on the flow from
a relativistic shock that breaks out of a star
\citep{nakayama05,pan06} show that significant acceleration also
occurs after the fluid is shocked. As the hot fluid expands
adiabatically, its thermal energy is converted to bulk kinetic energy.

Since all the above authors assume an ultrarelativistic equation of
state for the fluid, the final Lorentz factor their solutions predict
for the fluid is formally infinite as the fluid never cools.  They
avoid this difficulty by following fluid elements in the flow only to
the point where the fluid temperature becomes nonrelativistic and
approximating the final coasting Lorentz factor as the one given by
their solutions at that point. They thus find that the final Lorentz
factor of a given fluid element scales as $\gamma_0^{1+\sqrt{3}}$
where $\gamma_0$ is the Lorentz factor acquired by the fluid when it
is shocked. This method cannot accurately account for acceleration
that occurs around the time when the fluid cools to nonrelativistic
temperatures and can only produce approximate relations for the energy
and velocity of the ejecta. While \cite{kikuchi07} relax the
assumption of an ultrarelativistic fluid in their work on this
problem, they cannot completely characterize the acceleration while
the fluid is cooling either.

We approach this problem by introducing a new kind of self-similar
solution for the cooling and expanding fluid. In this solution, we
require that the fluid move at relativistic speeds but relax the
assumption that the fluid be hot.  We place the characteristic
position at the point where the fluid temperature is
transrelativistic. We thus exploit the self-similarity of the
transition between hot and cold fluid in the flow rather than the
self-similarity in the acceleration of the hot fluid. Indeed, this
flow when taken in its entirety is not self-similar: the size scales
that characterize the acceleration and the hot/cold transition evolve
with time according to different power laws. In other words, the
entire flow is a composite of two distinct self-similar solutions.  In
\S\ref{hotsolution} we summarize the solution for the hot fluid, which
gives the initial conditions for this new solution. In
\S\ref{coolingsolution} we derive the new solution, and in
\S\ref{compositesolution} we describe the behavior of fluid elements
in the composite solution. In \S\ref{specialk} we explain changes in
the flow's behavior for very shallow initial density profiles in the
stellar envelope. In \S\ref{finalgamma} we discuss the behavior of the
flow at late times and relate the elements' final Lorentz factors and
the Lorentz factors to which they were initially shocked. In
\S\ref{spherical} we find regions of the flow where spherical
corrections are important, and in \S\ref{summary} we summarize our
findings and discuss them in the context of previous work.  We take
the speed of light to be $c=1$ throughout our discussion.

\section{Initial conditions: behavior of the hot fluid} \label{hotsolution}

We are interested in the behavior at late times of a fluid flow which
begins as a relativistic shock propagating through the outer layers of
a star with a polytropic envelope. As long as the distance between the
front of the flow and the original location of the star's surface is
small compared to the star's radius, the geometry is planar. So we
seek a self-similar solution to the following hydrodynamic equations
representing energy, momentum, and mass conservation:
\begin{equation}
\label{e_conserve}
\frac{\partial}{\partial t}\left[\gamma^2(e+\beta^2 p)\right]
+ \frac{\partial}{\partial x}\left[\gamma^2\beta(e+p)\right]
 = 0
\end{equation}
\begin{equation}
\label{p_conserve}
\frac{\partial}{\partial t}\left[\gamma^2\beta(e+p)\right]
+ \frac{\partial}{\partial x}\left[\gamma^2(\beta^2 e+p)\right]
 = 0
\end{equation}
\begin{equation}
\label{n_conserve}
\frac{\partial}{\partial t}(\gamma n)
+ \frac{\partial}{\partial x}(\gamma\beta n)
 = 0 \;\;\; .
\end{equation}
The solution we seek must be connected to the hot flow, whose behavior
is well understood: \cite{sari05} and \cite{nakayama05} derive the
self-similar solution before the shock breaks out of the star and
\cite{pan06} derives the post-breakout solution.  Here we simply state
these results. We take $R(t)$ to be the characteristic position in the
solution, and
we set $R=0$ and $t=0$ at breakout.  We take $\Gamma$, $P$, and
$N$ to be the characteristic Lorentz factor, pressure, and number
density; $R$ and $\Gamma$ are related in that $\dot{R}=\sqrt{1-1/\Gamma^2}
\simeq 1-1/\left(2\Gamma^2\right)$. We define
\begin{equation}
\label{char_powerlaws_hot}
\frac{t\dot{\Gamma}}{\Gamma} = -\frac{m}{2} \;\;\; , \;\;\;
\frac{t\dot{P}}{P} = -m-k \;\;\; , \;\;\;
\frac{t\dot{N}}{N} = -\frac{m}{2} - k \;\;\; .
\end{equation}
Here $k$ gives the unshocked density profile in the stellar envelope:
assuming gravity is constant in the star's outermost layers and
unshocked pressure and density are related by a power law, the density
is given by a power law $\rho\propto |x|^{-k}$ where $x$ is the
position relative to the star's surface. This implies $-3\leq k\leq
-3/2$ for degenerate and convective envelopes and $k=-17/13$ for
Kramers opacity envelopes. We consider here the regime
$k<-\left(1+\sqrt{3}\right)/3$, which includes all of these profiles;
we explain this choice of maximum $k$ in \S\ref{coolingsolution}. We
write the solutions in the form
\begin{equation}
\label{gammadef_hot}
\gamma^2(x,t) = \frac{1}{2}\Gamma^2(t)g(\chi)
\end{equation}
\begin{equation}
\label{pdef_hot}
p(x,t) = P(t)f(\chi)
\end{equation}
\begin{equation}
\label{ndef_hot}
n(x,t) = N(t)\frac{h(\chi)}{g^{1/2}(\chi)}
\end{equation}
where the similarity variable is
\begin{equation}
\chi=1+2(m+1)\frac{R-x}{R/\Gamma^2} \;\;\; .
\end{equation}
Note that this expression is equivalent to 
\begin{equation}
\label{chi}
\chi = \frac{t-x}{t-R}
\end{equation}
taken in the limit where $\Gamma\gg 1$, or where $t\simeq
R(1+1/\left(2(m+1)\Gamma^2\right)$, so that the flow's characteristic
length scale is $t-R=R/\left(2(m+1)\Gamma^2\right)$.

Mathematically, the pre- and post-breakout solutions differ only in
the ranges in $\chi$ which apply. They are
\begin{eqnarray}
\label{chi_range}
-\infty < \chi < 1 \;\;\; , \;\;\; -\infty < g\chi < 1 & 
\;\;\;\;\;\;\;\;\; & t<0 \;\;\; \textrm{(pre-breakout)} \\
\infty > \chi > 0 \;\;\; , \;\;\; \infty > g\chi > \gchinought &
\;\;\;\;\; \;\;\;\; & t>0 \;\;\; \textrm{(post-breakout)}
\end{eqnarray}
where
\begin{equation}
\gchinought = 4+2\sqrt{3}-2k\sqrt{3} \;\;\; .
\end{equation}
These ranges differ because the sign of $R$ and the interpretation of
the characteristic values $\Gamma$, $P$, $N$, and $R$ change at
breakout. Before breakout, $R<0$ and $\Gamma$, $P$, $N$, and $R$ are
associated with the shock front; after breakout, $R>0$ and $\Gamma$,
$P$, $N$, and $R$ are associated with a fluid element which has
expanded by a factor of order unity.

The solutions are then completely specified by an expression for $m$ in
terms of $k$, which gives the time evolution of the flow, and expressions
for $g$, $f$, and $h$ in terms of $\chi$, which give the spatial profiles
for the hydrodynamic variables:
\begin{equation}
\label{m}
m=\left(3-2\sqrt{3}\right)k
\end{equation}
\begin{equation}
\label{g_hot}
g = \left| \frac{g\chi - \gchinought}{-1 + \gchinought}
    \right| ^{-\left(3-2\sqrt{3}\right)k}
\end{equation}
\begin{equation}
\label{f_hot}
f = \left| \frac{g\chi - \gchinought}{-1 + \gchinought}
    \right| ^{-\left(4-2\sqrt{3}\right)k}
\end{equation}
\begin{equation}
\label{h_hot}
h = \left| \frac{-g\chi + \gchinought}{1-\gchinought}
    \right|^{-\frac{\left( 2\sqrt{3}-3\right)(2k-1)k}
                   {-1+k\sqrt{3}-\sqrt{3}}}
    \left| g\chi-2 \right|^{\frac{k}{-1+k\sqrt{3}-\sqrt{3}}} \;\;\; .
\end{equation}
Thus defined, Eqs.~\ref{gammadef_hot}, \ref{pdef_hot}, \ref{ndef_hot}
satisfy the hydrodynamic equations with the equation of state
$p=e/3$. They are an accurate description of the flow with
$k<-\left(1+\sqrt{3}\right)/3$ only where the fluid is hot, or where $p/n\gg
1$. As the fluid expands and accelerates after breakout, it cools
adiabatically from the back of the flow towards the front
\citep{pan06}. As a result, the above post-breakout solution holds
only for a region at the front of the flow, and this region shrinks
with time.  This solution sets the boundary conditions for the new
solution we seek: as we approach the vacuum interface at $\chi=0$,
where the fluid is still hot, the new and the old solutions must
coincide.

\section{Self-similar solution for the cooling fluid} \label{coolingsolution}

We are interested in the behavior at late times of a fluid flow that
begins as a relativistic shock propagating through the outer layers of
a star with a polytropic envelope. We understand the behavior of the
part of the flow that is hot ($p/n\gg 1$) and therefore obeys the
equation of state $p=e/3$: it follows the self-similar solution given
in Eqs.~\ref{char_powerlaws_hot}, \ref{chi_range}--\ref{h_hot}.  We
note for convenience in our discussion below that the self-similar
variable $\chi$, as given by Eq.~\ref{chi}, is equivalent to
\begin{equation}
\label{chi_new}
\chi=\frac{t-x}{t-R}
\end{equation}
taken in the limit where $\Gamma\gg 1$, or where $t\simeq
R(1+1/(2(m+1)\Gamma^2)$. The implied characteristic length scale is
$t-R=R/(2(m+1)\Gamma^2)$.

As the fluid expands and accelerates after breakout, it cools
adiabatically from the back of the flow towards the front. At late
times, then, the post-breakout solution of \S 2, which we will refer
to here as the ``hot solution,'' holds only for a region at the very
front of the flow, and this region shrinks with time.  The hot
solution sets the boundary conditions for the new solution we seek: as
we approach the vacuum interface at $\chi=0$, the two solutions must
coincide.

In the new solution, which we will refer to as the ``cooling solution,''
we must include cold fluid. We therefore use the equation of state
\begin{equation}
\label{eos_cold}
p = \frac{1}{3}(e-n)
\end{equation} 
rather than the ultrarelativistic $p=e/3$. Although this equation of
state applies only to fluids with adiabatic index 4/3, our analysis
can be easily modified to accommodate an arbitrary equation of state
of the form
\begin{equation}
\label{eos_general}
\frac{p}{n} = F\left(\frac{e}{n}\right)
\end{equation}
where $F$ is an invertible function since the hydrodynamic equations
can be written in self-similar form with any such equation of state.

We must also specify a characteristic scale and define the
characteristic Lorentz factor, pressure, and number density to be
consistent with this scale.  We seek the profiles of the hydrodynamic
variables in the region where the fluid temperature transitions from
hot to cold; the natural scale for this transition is the distance
$\delta$ between the vacuum interface, where the fluid is hottest, and
the point where the fluid temperature becomes nonrelativistic. We set
this point to be where $p/n=1$. Then the similarity variable is
\begin{equation}
\xi=\frac{t-x}{\delta}
\end{equation}
and, by analogy with the hot solution, we express
$\gamma$, $p$, and $n$ as
\begin{equation}
\label{gammadef_cold}
\gamma^2(x,t) = \frac{1}{2}\Gammabar^2(t)\gbar(\xi)
\end{equation}
\begin{equation}
\label{pdef_cold}
p(x,t) = \Pbar(t)\fbar(\xi)
\end{equation}
\begin{equation}
\label{ndef_cold}
n(x,t) = \Nbar(t)\frac{\hbar(\xi)}{\gbar^{1/2}(\xi)} \;\;\; .
\end{equation}
We choose $\Gammabar$, $\Pbar$, and $\Nbar$, the new characteristic
values of the Lorentz factor, pressure, and number density, to match
the $\gamma$, $p$, and $n$ values given by the hot solution where
$p/n=1$. We take $\chicold$ to be the value of the old similarity
variable $\chi$ corresponding to $p/n=1$ in the hot solution:
\begin{equation}
\label{gammabar_def}
\Gammabar^2 = \Gamma^2 g(\chicold)
\end{equation}
\begin{equation}
\label{pbar_def}
\Pbar = P f(\chicold)
\end{equation}
\begin{equation}
\label{nbar_def}
\Nbar = N \frac{h(\chicold)}{g^{1/2}(\chicold)} \;\;\; .
\end{equation}
This choice of characteristic values dictates
\begin{equation}
\label{pbar_eq_nbar}
\Pbar=\Nbar \;\;\; .
\end{equation}
In the limit of late times, when $\delta \ll R/\Gamma^2$ and
$g\chi-\gchinought \ll \gchinought$, Eqs.~\ref{gammabar_def} and \ref{chi_new}
give
\begin{equation}
\gchinought \frac{\Gamma^2}{\Gammabar^2} 
 \simeq \chicold 
 \simeq \frac{\delta}{t-R}
 \simeq \delta\cdot\frac{2(m+1)\Gamma^2}{t}
\end{equation}
\begin{equation}
\label{delta}
\delta
 = \frac{\gchinought}{2(m+1)}\frac{t}{\Gammabar^2} 
 = \left(2+\sqrt{3}\right)\frac{t}{\Gammabar^2} \;\;\; .
\end{equation}
Note that the characteristic scale $t-R\sim R/\Gamma^2$ in the
post-breakout solution for the hot fluid is different from the new scale
$\delta$. $R$ is the location of a fluid element that has expanded by
a factor of order unity since breakout. Because $R$ evolves according
to the finite characteristic Lorentz factor $\Gamma$, $R$ lags
farther and farther behind the front of the flow, where the Lorentz
factors are arbitrarily large. Since $\Gamma$ decreases with time as
per Eq.~\ref{char_powerlaws_hot}, $t-R$ increases with time. In the limit
of late times, then, $R$ lags far behind the portion of the flow where
the fluid remains hot and $t-R$ becomes much larger than the space
occupied by the hot fluid. In other words, the scale $t-R$ that
characterizes the hot solution becomes irrelevant to the transition
between hot and cold fluid that is of interest here.

To get $\Gammabar$ and $\delta$ as functions of time, we apply
Eqs.~\ref{g_hot}--\ref{h_hot} at the point $p/n=1$. 
We use $P/N=\Gamma/(3\sqrt{2})$, a relation that follows from
the shock jump conditions applied in the pre-breakout solution.
\begin{eqnarray}
\label{where_cold}
1&
     =& \frac{p}{n}
     =  \frac{\Gamma}{3\sqrt{2}}
        \frac{\sqrt{g(\chicold)}f(\chicold)}
             {h(\chicold)}\\
&
     =& \frac{\Gamma}{3\sqrt{2}}
        \left[g(\chicold)\right]
             ^{\frac{1}{m}
               \left(\frac{3m}{2}+k+\frac{m(2k-1)}
                                         {-1-\sqrt{3}+k\sqrt{3}}\right)}
        \left[g(\chicold)\cdot \chicold-2
        \right]^{\frac{-k}{-1-\sqrt{3}+k\sqrt{3}}}
\end{eqnarray}
\begin{eqnarray}
\label{gammabar}
\Gammabar^2&
     =& \Gamma^2 g(\chicold) \\
&
     =& \Gamma^2
        \left(\frac{3\sqrt{2}}{\Gamma} 
              \left(g(\chicold)\cdot \chicold-2
              \right)^{\frac{k}{-1-\sqrt{3}+k\sqrt{3}}}
        \right)^{m/\left(\frac{m}{2}-\frac{k}{-1-\sqrt{3}+k\sqrt{3}}\right)} 
\;\;\; .
\end{eqnarray}
This gives
\begin{equation}
\label{gammabardot}
a
 = \frac{t\dot{\Gammabar}}{\Gammabar}
 = \frac{k\sqrt{3}}{1+\sqrt{3}+3k}
\end{equation}
\begin{equation}
\label{deltadot}
\frac{t\dot{\delta}}{\delta}
 = \frac{1+\sqrt{3} + \left(3-2\sqrt{3}\right)k}{1+\sqrt{3}+3k}
\end{equation}
at late times, when $g(\chicold)\cdot\chicold\simeq \gchinought$.
Solving Eq.~\ref{where_cold} for $f(\chicold)$ or
$h(\chicold)/\sqrt{g(\chicold)}$ similarly gives
\begin{equation}
\label{pbardot}
b
 = \frac{t\dot{\bar{P}}}{\Pbar}=\frac{t\dot{\Nbar}}{\Nbar}
 = -\frac{4k}{1+\sqrt{3}+3k} \;\;\; .
\end{equation}
Note that $k=-\left(1+\sqrt{3}\right)/3\simeq -0.91$ makes $a$
and $b$ diverge in Eqs.~\ref{gammabardot} and \ref{pbardot}. This $k$
marks a qualitative change in the behavior of the flow which we
discuss further in \S~\ref{specialk}.

\begin{figure}
\plotone{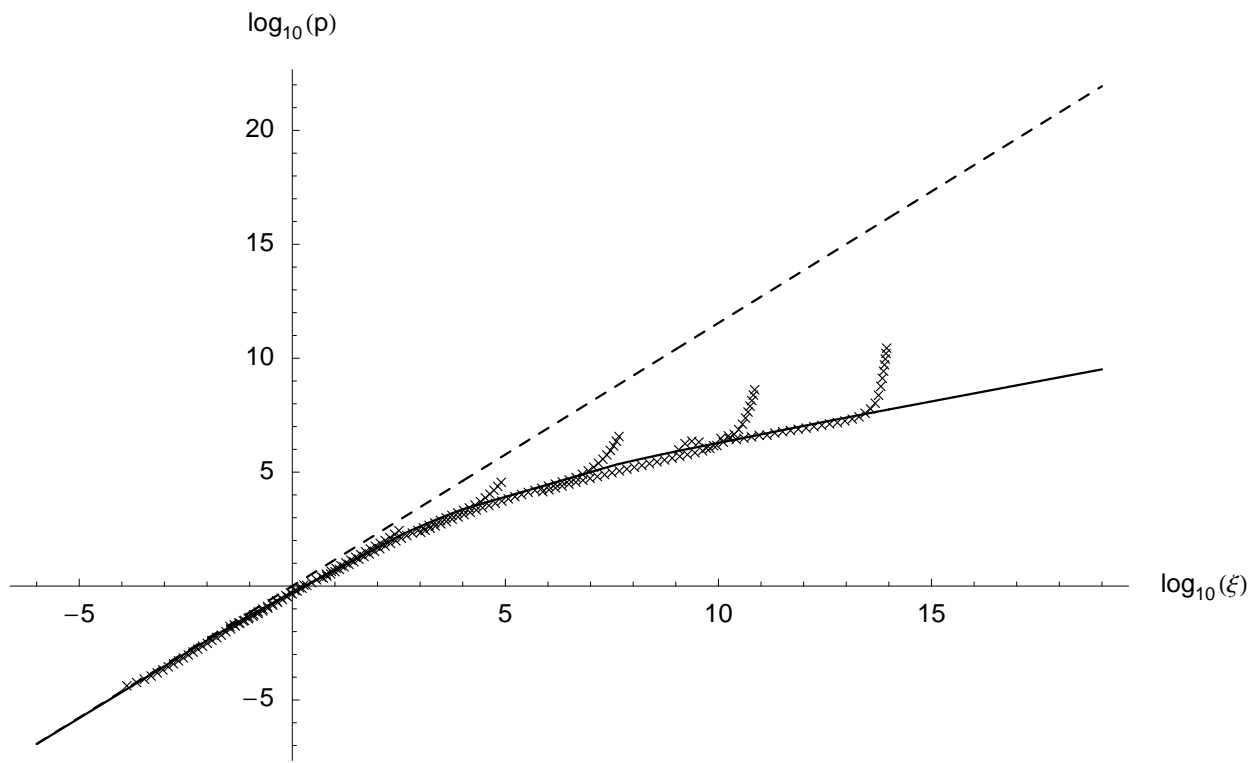}
\caption{Profile of the pressure $p$ as a function of the similarity
  variable $\xi$ for $k=-3$. The dashed line is the hot solution valid
  for the fluid near the front, at small $\xi$; the solid line is the
  cooling solution. Data from numerical simulations are shown as
  crosses. In order to cover a substantial range in $\xi$, data from
  six $p$ vs. $\xi$ profiles corresponding to different times in
  the same simulation run are shown. The data agree well with the
  cooling solution. The ``tails'' at the ends of the numerical
  simulation profile data are due to edge effects at the ends of the
  simulation grid that are not self-similar. The overall $y$-axis
  normalization is arbitrary, but the relative normalizations of the
  hot solution, the cooling solution, and the numerical simulations
  are correct. }
\label{p_profiles}
\end{figure}

\begin{figure}
\plotone{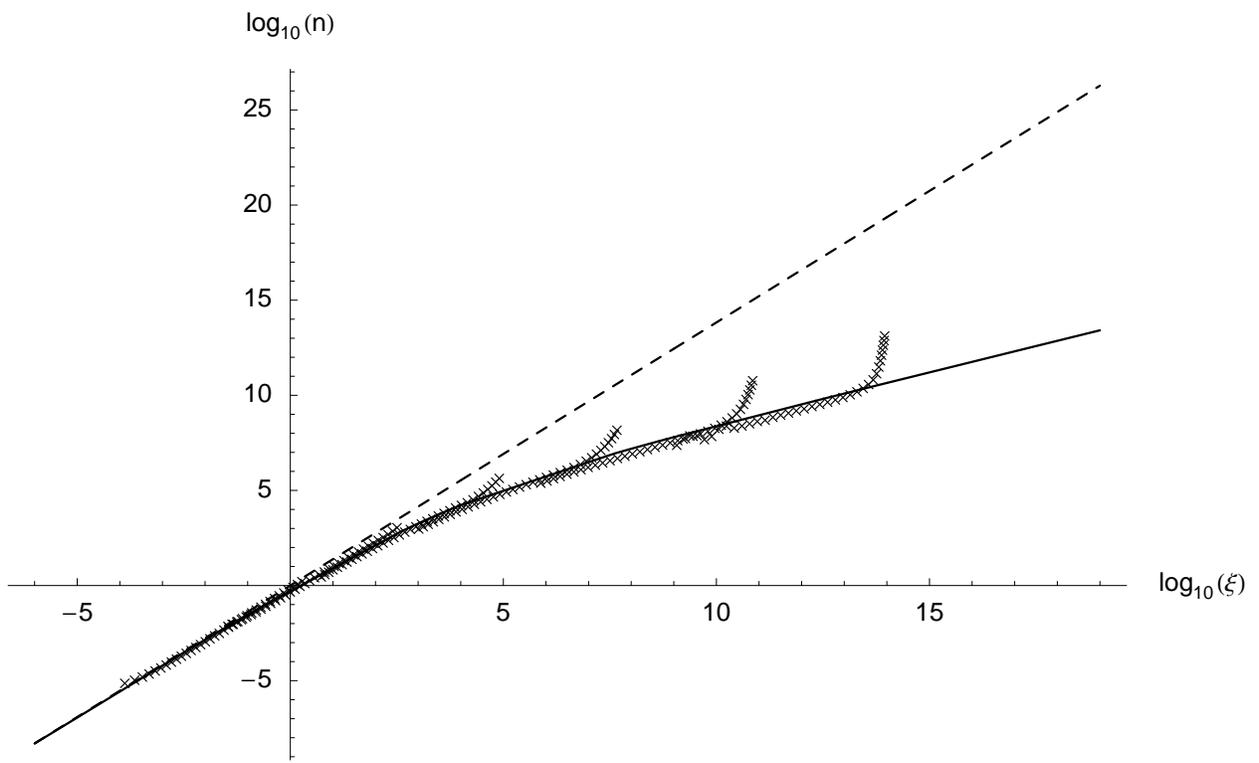}
\caption{Same as Figure~\ref{p_profiles} for the number density $n$ rather
than $p$.}
\label{n_profiles}
\end{figure}

\begin{figure}
\plotone{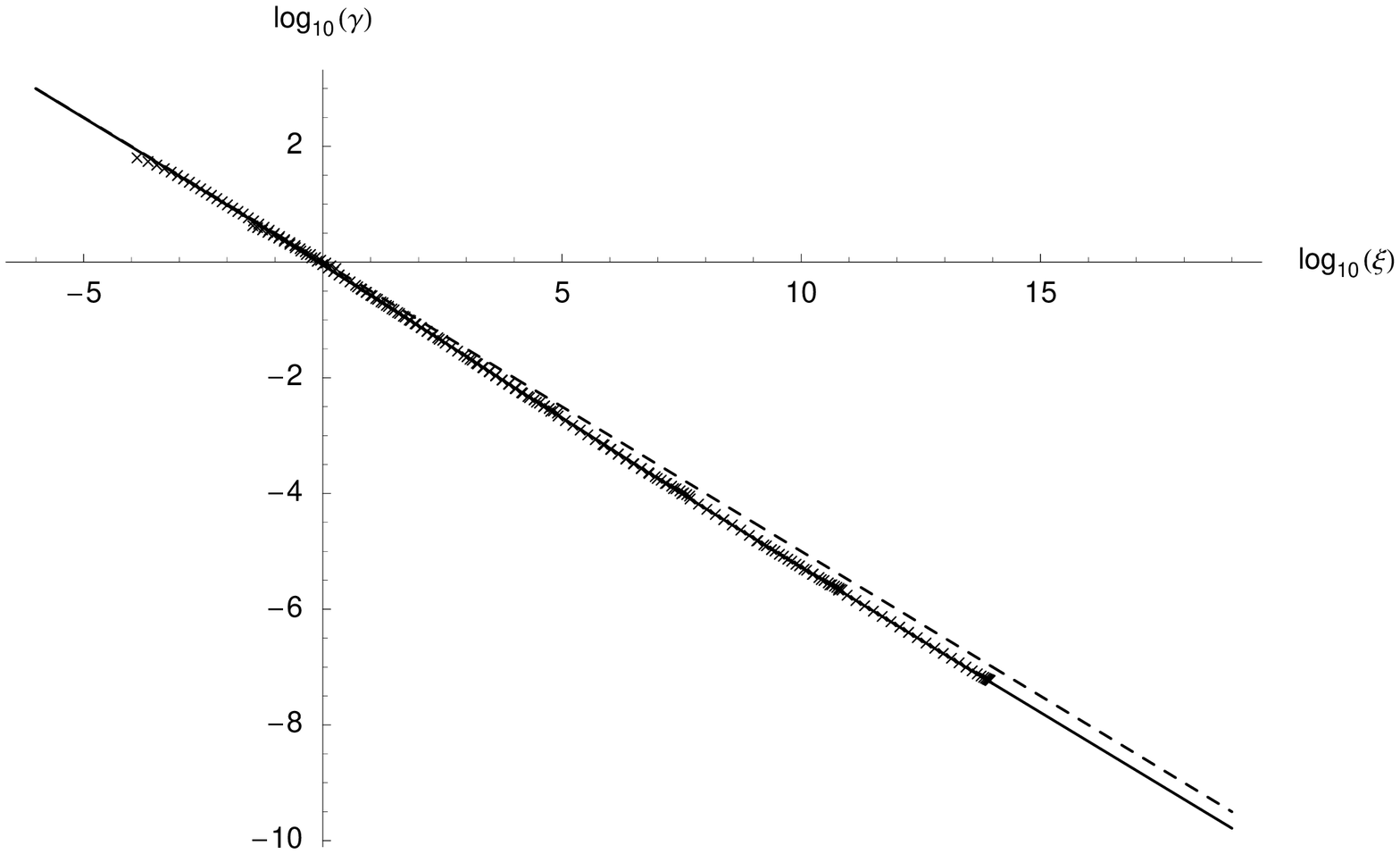}
\caption{Same as Figure~\ref{p_profiles} for the Lorentz factor $\gamma$
rather than $p$.}
\label{g_profiles}
\end{figure}

We now proceed to solve the hydrodynamic equations for the relevant
range in $k$. We use the equation of state Eq.~\ref{eos_cold} to
rewrite Eq.~\ref{e_conserve}, Eq.~\ref{n_conserve}, and the difference
equation obtained by subtracting Eq.~\ref{p_conserve} from
Eq.~\ref{e_conserve}.  We take the limit $\gamma \gg 1$. We rewrite
the differentiation operators as
\begin{equation}
\frac{\partial}{\partial t}
 = \dot{\Gammabar}\frac{\partial}{\partial\Gammabar}
   + \dot{P}\frac{\partial}{\partial P}
   + \dot{N}\frac{\partial}{\partial N}
   + \frac{1}{\delta}\left(1-\xi\dot{\delta}\right)\frac{\partial}{\partial\xi}
\end{equation}
\begin{equation}
\frac{\partial}{\partial x}
 = -\frac{1}{\delta}\frac{\partial}{\partial\xi}
\end{equation}
and substitute these and Eqs.~\ref{gammabar_def}-\ref{pbar_eq_nbar},
\ref{gammabardot}, and \ref{pbardot} to get
\begin{eqnarray}
\label{diffeq1}
\nonumber
0&
 =& b\left(2\fbar+\frac{\hbar}{\gbar^{1/2}}\right)
    + \frac{t}{\delta\Gammabar^2}
      \left[-\gbar'\left(4\frac{\fbar}{\gbar^2}
                         +\frac{3}{2}\frac{\hbar}{\gbar^{5/2}}\right)
            +\frac{4\fbar'}{\gbar}
            +\frac{\hbar'}{\gbar^{3/2}}
      \right] \\
&
&   \;\;-\: \xi\frac{t\dot{\delta}}{\delta}
      \left[-\frac{\gbar'}{2}\frac{\hbar}{\gbar^{3/2}}
            +2\fbar'
            +\frac{\hbar'}{\gbar^{1/2}}
      \right]
\end{eqnarray}
\begin{eqnarray}
\label{diffeq2}
\nonumber
0&
 =& (2a+b)\left(2\gbar\fbar+\frac{\hbar\gbar^{1/2}}{2}\right)
    + \frac{t}{\delta\Gammabar^2}
      \left[-\frac{\gbar'}{4}\frac{\hbar}{\gbar^{3/2}}
            +\fbar'
            +\frac{\hbar'}{2\gbar^{1/2}}\right] \\
&
&   \;\;-\: \xi\frac{t\dot{\delta}}{\delta}
      \left[\gbar'\left(2\fbar+\frac{\hbar}{4\gbar^{1/2}}\right)
            +2\fbar'\gbar
            +\frac{\hbar'}{2}\gbar^{1/2}\right]
\end{eqnarray}
\begin{equation}
\label{diffeq3}
0
 = (a+b)\hbar 
   + \frac{t}{\delta\Gammabar^2}
     \left[-\gbar'\frac{\hbar}{\gbar^2}+\frac{\hbar'}{\gbar}\right]
   - \xi\frac{t\dot{\delta}}{\delta} \hbar' \;\;\; .
\end{equation}
We substitute Eqs.~\ref{delta} and \ref{deltadot} into
Eqs.~\ref{diffeq1}, \ref{diffeq2}, and \ref{diffeq3} and integrate
this ODE system numerically to produce the solution shown in
Figures~\ref{p_profiles}, \ref{n_profiles}, and
\ref{g_profiles}. These figures also compare our solution to a
one-dimensional numerical simulation of a relativistic planar shock
wave which accelerates through and breaks out of a $k=-3$ medium and
then expands and cools through the transrelativistic regime. The
initial conditions used in the simulation represent the shock as an
unresolved contact discontinuity with $\Gamma=5$ at $x=-10^{-5}$ at
starting time $t=-1.0\times 10^{-5}$. The simulation grid contains 400
fluid elements spread logarithmically over two orders of magnitude in $x$;
our calculation follows these elements until time $t=2.6\times 10^8$.

\begin{figure}
\plotone{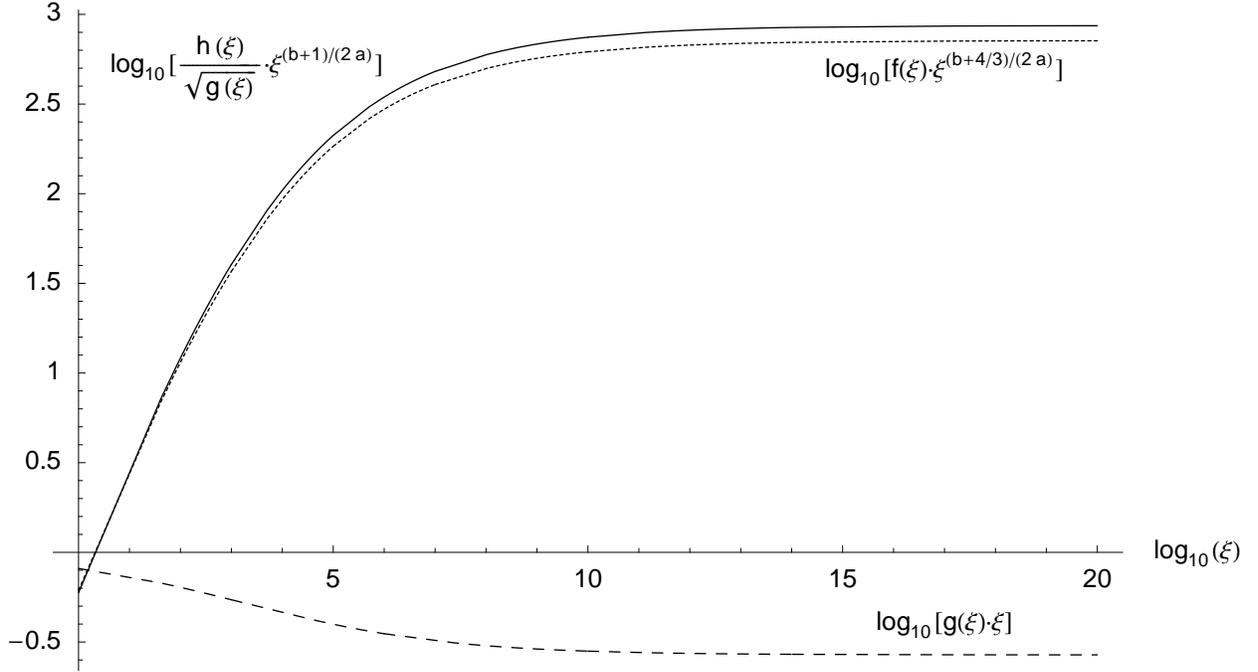}
\caption{Verification of the scalings of $\gbar$, $\fbar$, and $\hbar$
  with $\xi$ at large $\xi$, or cold fluid temperatures. The functions
  plotted (dashed line for $\gbar$, dotted line for $\fbar$, solid
  line for $\hbar$) were obtained via numerical integration of the
  ODEs in Eqs.~\ref{diffeq1}, \ref{diffeq2}, \ref{diffeq3}. They show
  that Eqs. \ref{gbar_scaling}, \ref{p_scaling},
  \ref{n_scaling}---relations derived for fluid which has finished
  accelerating---are valid at large $\xi$. In particular, $\gbar\xi$
  approaches the expected value $2-\sqrt{3}=10^{-0.572}$. In this
  calculation we used $k=-3$.}
\label{gfh_scalings}
\end{figure}

We can check that the behavior of this solution at large $\xi$---where
the fluid is very cold and where the hot solution and cooling solution
differ most---is physical. Consider a fluid element many distance
scales $\delta$ behind the vacuum interface at position $t-x\gg
\delta$.  This fluid element must have become cold at some time
$t_\mathrm{cold}\ll t$; as a result, it has long since stopped
accelerating and has spent most of the time interval
$t-t_\mathrm{cold}$ coasting at its current Lorentz factor $\gamma$.
Then this fluid element has
\begin{equation}
\label{xi_coasting}
\xi = \frac{t-x}{\delta} \simeq \frac{t-(t-t_\mathrm{cold})\sqrt{1-1/\gamma^2}}{\delta} \simeq \frac{t/\left(\Gammabar^2\gbar\right)}{t/\Gammabar^2\cdot\left(2+\sqrt{3}\right)} = \frac{2-\sqrt{3}}{\gbar}
\end{equation}
so at large $\xi$ we expect
\begin{equation}
\label{gbar_scaling}
\gbar\xi=2-\sqrt{3} \;\;\; .
\end{equation}
We cannot get exact relations for $\fbar$ and $\hbar$ in the 
large $\xi$ limit in this way because $p$ and $n$ change significantly
while the fluid element finishes its acceleration. However, we can
check the scalings of $\fbar$ and $\hbar$ with $\xi$. Because the fluid 
elements far behind the front are coasting with Lorentz factors that
are virtually constant in time, the volume of each fluid element increases
linearly with time. This implies 
\begin{equation}
n\propto t^{-1} \;\;\; , \;\;\; p\propto n^{4/3}\propto t^{-4/3}
\end{equation}
for a single fluid element. From the definitions of $\Gammabar$,
$\Pbar$, and $\Nbar$ in Eqs.~\ref{gammabar_def}, \ref{pbar_def}, and
\ref{nbar_def}, we know $\gamma(\xi=1)/\Gammabar$, $p(\xi=1)/\Pbar$,
and $n(\xi=1)/\Nbar$ are constant in time. Then for a single fluid
element,
\begin{equation}
p\sim p(t_\mathrm{cold})\left(\frac{t}{t_\mathrm{cold}}\right)^{-4/3}
\propto t_\mathrm{cold}^{b+4/3} \;\;\; .
\end{equation}
Since 
\begin{equation}
\gbar= \frac{2\gamma^2}{\Gammabar^2(t)}\propto \frac{\Gammabar^2(t_\mathrm{cold})}{\Gammabar^2(t)}\propto t_\mathrm{cold}^{2a} \;\;\; ,
\end{equation}
we have 
\begin{equation}
\label{p_scaling}
\xi\propto t_\mathrm{cold}^{-2a}\propto p^{-\frac{2a}{b+4/3}}
\;\;\; \longrightarrow \;\;\; 
p\propto \xi^{-\frac{b+4/3}{2a}} \;\;\; .
\end{equation}
A similar calculation yields
\begin{equation}
\label{n_scaling}
n\propto \xi^{-\frac{b+1}{2a}} \;\;\; .
\end{equation}
That the relations in Eqs.~\ref{gbar_scaling}, \ref{p_scaling}, and
\ref{n_scaling} hold at large $\xi$ is shown in Figure~\ref{gfh_scalings}.

\section{The composite solution} \label{compositesolution}

As discussed in \S~\ref{hotsolution} and \ref{coolingsolution}, the
hot fluid close to the vacuum interface and the cooling fluid further
back in the flow obey two different self-similar solutions in which
both the time evolution and the physical interpretation of the
characteristic length scale, Lorentz factor, pressure, and density
differ. In other words, we describe the entire post-breakout
flow---which, when taken as a whole, is not self-similar---by a
composite of two self-similar solutions built up around the two
different length scales which characterize different portions of the
flow. This to our knowledge is the first such `composite solution'
found.  \cite{chevalier82}, for example, also uses two self-similar
solutions in a description of a single flow---specifically, the
interaction between an expanding shell of fluid and a stationary
external medium.  However, his solutions have the same characteristic
length scale and time evolution; they differ only in the shapes of
their profiles and in the disjoint regions of the flow in which they
operate. They may be considered as a single self-similar solution in
which the pressure, velocity, and density profiles are piecewise
functions of position.

We can check that the behaviors of fluid elements and sound waves in
the hot solution and the cooling solution are consistent.  The
characteristic position $R$ in the hot solution moves backwards
relative to the vacuum interface with time because $\Gamma$ decreases
with time. By contrast, the characteristic position $t-\delta$ in the
cooling solution moves forwards relative to $ct$ because fluid
elements at the back of the solution cool faster than those at the
front, and $t-\delta$ marks the location of a fluid element that has
just cooled.  We confirm the forward motion by looking at
Eq.~\ref{deltadot}, which indeed gives $t\dot{\delta}/\delta<0$ for
the range of $k$ of interest ($k<-\left(1+\sqrt{3}\right)/3$).  We expect fluid
elements in the cooling solution to move backwards in the solution, or
towards larger $\xi$: every fluid element must eventually finish
accelerating and become cold, so the point $t-\delta$ that marks the
hot/cold transition must overtake every fluid element. Indeed, the
time derivative of $\xi$ following a fluid element,
\begin{equation}
\label{DxiDt}
\frac{D\xi}{Dt}
 = \left(\frac{d}{dt} + \sqrt{1-\frac{1}{\gamma^2}}\frac{d}{dx}\right)\xi
 = \frac{\xi}{t}
   \left(\frac{2-\sqrt{3}}{\gbar\xi}-\frac{t\dot\delta}{\delta}\right) \;\;\; ,
\end{equation}
is always positive since $t\dot{\delta}/\delta<0$.  Fluid elements in
the hot solution move forwards with time, towards smaller $g\chi$, since
they accelerate while $R$ decelerates. The advective time derivative
of $\chi$,
\begin{equation}
\label{DchiDt}
\frac{D\chi}{Dt}
 = \frac{\chi}{t}\left(\frac{2}{g\chi}-1\right)(m+1) \;\;\; ,
\end{equation}
is always negative since $g\chi\geq\gchinought > 4+2\sqrt{3}$ everywhere.

If the proper sound speed in the fluid is
$\beta_s=\sqrt{4/3}\fbar^{1/2}\gbar^{1/4}\hbar^{-1/2}$, then the motion of a
sound wave in the cooling solution is given by
\begin{equation}
\frac{d\ln\xi_\pm}{d\ln t}
 = \frac{t}{\xi_\pm}
   \frac{1}{\delta}\left(1-\frac{dx_\pm}{dt}-\xi_\pm \dot{\delta}\right)
 = \frac{2-\sqrt{3}}{\gbar\xi}\left(\frac{1\mp\beta_s}{1\pm\beta_s}\right)
   - \frac{t\dot{\delta}}{\delta}
\end{equation}
where the signs denote forward- and backward-propagating sound waves.
Again, $t\dot{\delta}/\delta<0$, so $d\xi_\pm/dt>0$ everywhere and all
sound waves move backwards in the cooling solution. In other words, all
fluid elements are disconnected from the vacuum interface. In the
sense that the front is disconnected from the fluid far back in the
flow, the cooling solution is similar to Type II solutions. However, in
contrast to the usual Type II scenario, there is no sonic point
constraining the solution.

\begin{figure}
\plotone{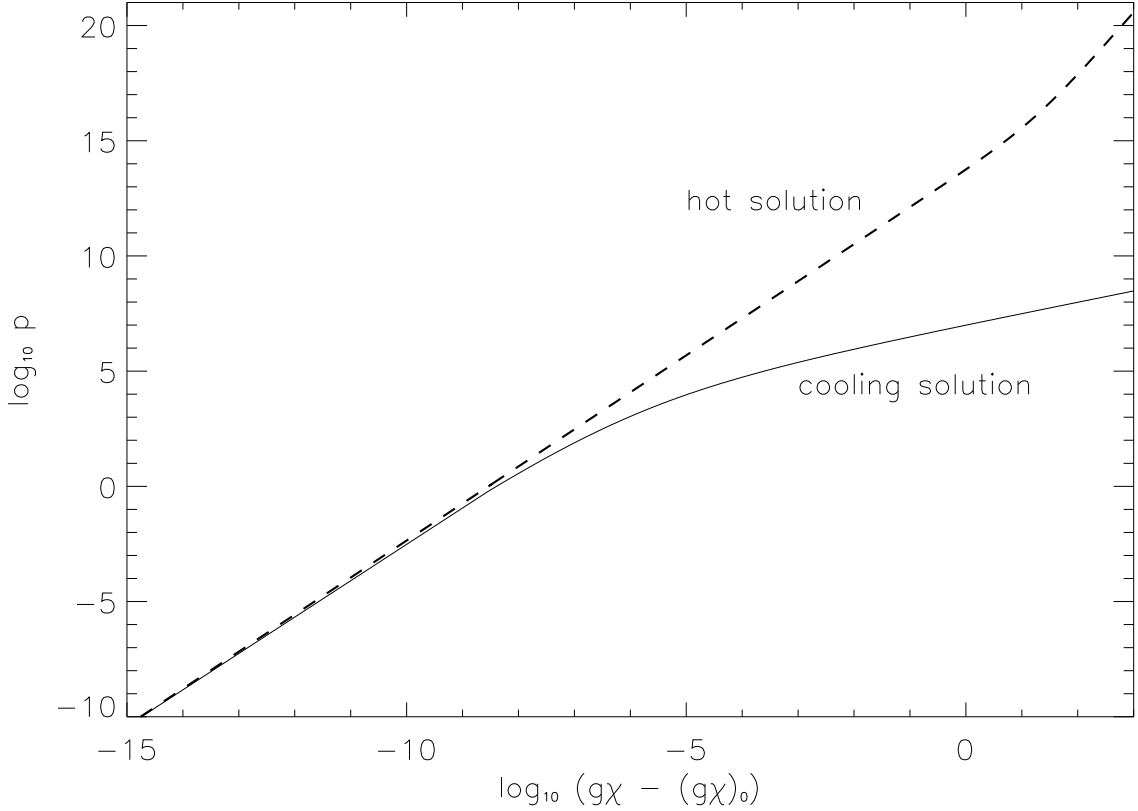}
\caption{Pressure profiles for initial density profile $k=-3$ at late
  times when the cooling solution applies. The dashed curve is the hot
  solution and the solid curve is the cooling solution. The front of
  the flow, where $g\chi\simeq \gchinought$, is toward the left. As in
  Figure~\ref{p_profiles}, the overall $y$-axis normalization is
  arbitrary but the relative normalizations of the hot and cooling
  solutions are correct. The `kinks' in the hot and cooling solution
  curves occur at their respective characteristic scales $t-R$ and
  $\delta$. Fluid far back in the flow, where $\xi\gg 1$ or,
  equivalently, $g\chi-\gchinought\gg \gchicold-\gchinought$, follows
  the cooling solution. This cold fluid falls behind the position
  predicted for it in the hot solution because it is no longer
  accelerating. As a result, the hot solution gives artificially high
  pressures for fluid at the back of the flow. The hot solution
  applies only at the front of the flow in a region small compared to
  both characteristic scales. As this region corresponds to the limit
  in which we set the cooling solution to match the hot solution, the
  entire flow may be described with the cooling solution alone at
  these times when $\Gamma<1$.}
\label{schematic_k300}
\end{figure}

There is a caveat in this composite view of the flow for the density
profiles $k<-\left(1+\sqrt{3}\right)/3$ discussed in the preceding
sections.  At early times when $\Gamma\gg 1$, all the fluid elements
which have accelerated by at least a factor of order unity are those
with $p/n\gtrsim\Gamma$; slower and cooler fluid elements still have
very nearly the same temperatures and speeds as they had just after
being shocked. So the only fluid elements with $p/n$ of order unity or
smaller are those which were never shocked to relativistic
temperatures, and the part of the flow moving relativistically can be
described with the hot solution alone.  At late times when the cold
solution is relevant, the hot solution applies only in a region at the
front of the flow whose size is much smaller than the characteristic
scale $R/\Gamma^2$. In this small region, the profiles of the
hydrodynamic variables in hot solution appear very nearly scale free,
and the cooling solution approaches the resulting power-law profiles
toward the front of the flow, in the limit of small $\xi$. Then the
hot solution is unnecessary to the description of the flow at these
late times, so we can think of the flow as following the hot solution
at early times and transitioning to the cooling solution when
$\Gamma=1$. Figure~\ref{schematic_k300} is a schematic of the composite
flow at these late times.

However, a composite of the hot and cooling solutions is essential in
describing the flow for density profiles
$-\left(1+\sqrt{3}\right)/3\leq k<0$, which we discuss in the next
section.

\section{Solutions when {$-\left(1+\sqrt{3}\right)/3\leq \MakeLowercase{k}<0$}} \label{specialk}

We restricted our discussion of the hot and cooling solutions in
\S~\ref{hotsolution}, \ref{coolingsolution}, \ref{compositesolution}
to initial density profiles $k<-\left(1+\sqrt{3}\right)/3$. That the
initial density decrease toward the star's surface requires $k<0$, but
so far we have neglected the interval $-\left(1+\sqrt{3}\right)/3\leq
k<0$. To describe the flow for these $k$, we return to the singularity
at $k=-\left(1+\sqrt{3}\right)/3$ noted in \S~\ref{coolingsolution}
and examine the behavior of $\gchicold=g(\chicold)\cdot\chicold$.

Eq.~\ref{where_cold} implies
\begin{equation}
\label{dlngchicold}
\frac{d\ln \gchicold}{d\ln t}
   = -\frac{\gchicold-(g\chi)_0}{\gchicold} 
      \cdot \frac{\gchicold-2}{\gchicold-2\left(3+4/\sqrt{3}\right)} \;\;\; .
\end{equation}
For $-\left(1+\sqrt{3}\right)/3<k<0$, integration of
Eq.~\ref{dlngchicold} shows that $\gchicold$ is a double-valued
function of time until the entire flow cools. As $t\rightarrow 0$ from
above, $\gchicold\rightarrow \{\gchinought,\infty\}$; for positive
times, the smaller value of $\gchicold$ increases and the larger value
decreases with time until $\gchicold$ becomes single-valued at
$\gchicold=2\left(3+4/\sqrt{3}\right)$, which occurs at a finite
time. In other words, fluid cools quickly both at the front, where the
expansion timescale is shortest, and at the very back, where the
shocked fluid is coldest. 

\begin{figure}
\plotone{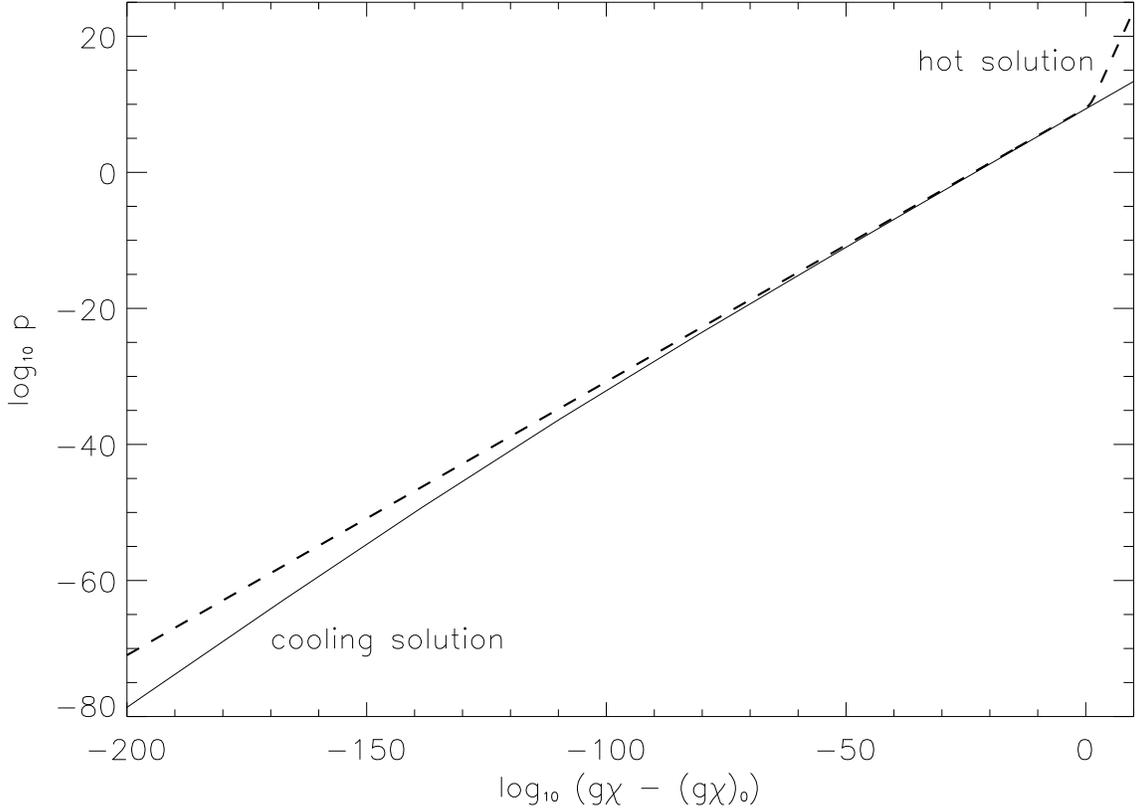}
\caption{Same as Figure~\ref{schematic_k300}, but for initial density
  profile $k=-3/4$ at early times, before $\gchicold$ becomes
  single-valued. Here the cooling solution applies at the front of the
  flow; the hot solution applies further back in the flow. Because the
  cold fluid has stopped accelerating, the hot solution gives
  artificially high pressures for fluid towards the front of the flow
  where $\xi\ll 1$ or
  $g\chi-\gchinought\ll\gchicold-\gchinought$. However, the cooling
  solution misses the transition between regions of hot fluid which
  have and have not accelerated significantly since being shocked;
  this transition occurs at $\chi\sim 1$, or $g\chi-\gchinought$ of order
  unity. An accurate description of the entire flow requires both the
  hot and cooling solutions.}
\label{schematic_k075}
\end{figure}

Since $\gchicold$ becomes single-valued at a $\gchicold$ of order
unity, the value of $g$ corresponding to the larger value of
$\gchicold$ is always of order unity or smaller. But $g=1$ for a fluid
element which has accelerated by a factor of order unity, so this
larger $\gchicold$ tracks fluid elements which cool before or just as
they manage to accelerate by a factor of order unity. These fluid
elements cannot have been shocked to relativistic temperatures, so we
focus on the evolution of the smaller value of $\gchicold$ as it moves
through the front of the solution. In contrast to the
$k<-\left(1+\sqrt{3}\right)/3$ case, the fluid near the front cools from the
vacuum interface backwards, from the fastest-moving toward
slower-moving fluid, and the fluid farthest forward is coldest:
$\Gammabar$ decreases with time and $\Pbar$, $\Nbar$, $\delta$
increase with time. Fluid elements just behind the
front---specifically, those with $g\chi-\gchinought\ll
2+2/\sqrt{3}+2k\sqrt{3}$ ---always have $\gchicold\simeq\gchinought$,
so they obey Eqs.~\ref{gammabardot}-\ref{pbardot}. Then the discussion
of the cooling solution in \S~\ref{coolingsolution} applies in the
limit of small positive times, when the front of the flow is cooling,
except in that the initial conditions used to integrate
Eqs.~\ref{diffeq1}-\ref{diffeq3} are set by matching to the hot
solution far back in the cooling flow, at $\xi\gg 1$, rather than at
the vacuum interface. Between breakout and the time when all the
fluid cools, the flow contains both hot and cold fluid moving at
relativistic bulk speeds; it therefore follows a true composite of the
hot and cooling solutions. Figure~\ref{schematic_k075} shows a
schematic of this composite solution.

As in the $k<-\left(1+\sqrt{3}\right)/3$ case, fluid elements move
forwards in the hot solution with time according to Eq.~\ref{DchiDt}.
No qualitative change in the time evolution of the scale $R$ ---or,
therefore, in the behavior of fluid in the hot solution---occurs as
$k$ increases through $-\left(1+\sqrt{3}\right)/3$. In contrast to the
$k<-\left(1+\sqrt{3}\right)/3$ case, fluid elements move forwards in
the cooling solution as well. Eq.~\ref{xi_coasting}, which still holds
in the limit $t-x\ll\delta$, implies $\gbar\xi\geq 2-\sqrt{3}$, while
Eq.~\ref{deltadot} implies $t\dot{\delta}/\delta >1$. Then from
Eq.~\ref{DxiDt} we have $D\xi/Dt <0$.

For $k=-\left(1+\sqrt{3}\right)/3$,
$\gchinought=3\left(2+4/\sqrt{3}\right)$, so $\gchicold$ is
single-valued and $\gchicold-2\propto t^{-1}$. Then there is a finite
time at which $\gchicold$ reaches $\gchinought$, $\delta$ shrinks to
0, and the entire flow cools. Also, at any $t>0$, Eq.~\ref{f_hot},
\ref{h_hot} imply that the temperature at the front of the flow, where
$g\chi-\gchinought\ll \gchinought$, is roughly independent of
position, and the temperature at the vacuum interface is finite; all
the fluid at the front cools at the same rate. Though
Eq.~\ref{deltadot} does not work for this $k$, we know $\delta$
decreases monotonically to 0 with time, so Eq.~\ref{DxiDt} gives
$D\xi/Dt>0$. So fluid elements move backwards in the cooling solution
until $\gchicold=\gchinought$ and $\delta=0$.

\section{Behavior of fluid elements at late times} \label{finalgamma}

Earlier analytic work has established that the final Lorentz factor
$\gamma_\mathrm{final}$ of a given fluid element should scale
according to
\begin{equation}
\label{final_gamma_propto}
\gamma_\mathrm{final}=K \gamma_\mathrm{shocked}^{1+\sqrt{3}}
\end{equation}
where $\gamma_\mathrm{shocked}$ is the fluid element's Lorentz factor
immediately after it is shocked in the pre-breakout flow and the
coefficient $K$ is independent of $\gamma_\mathrm{shocked}$
\citep{johnson71,pan06}.  \cite{tan01} have found numerically that
$K\simeq 2.6$ for $k=-3$. They note, and we confirm from our own
experience, that it is difficult to continue numerical simulations
until the very end of the fluid acceleration since the conversion of
thermal to bulk kinetic energy is quite slow:
Figure~\ref{gfh_scalings} indicates significant acceleration until
$\xi\sim 10^{10}$. To estimate the coefficient, \cite{tan01}
applied correction factors to their simulation results of up to
$\sim$50\% for fluid elements with final Lorentz factors of order
$\sim 10^3$.

\begin{figure}
\epsscale{1}
\plotone{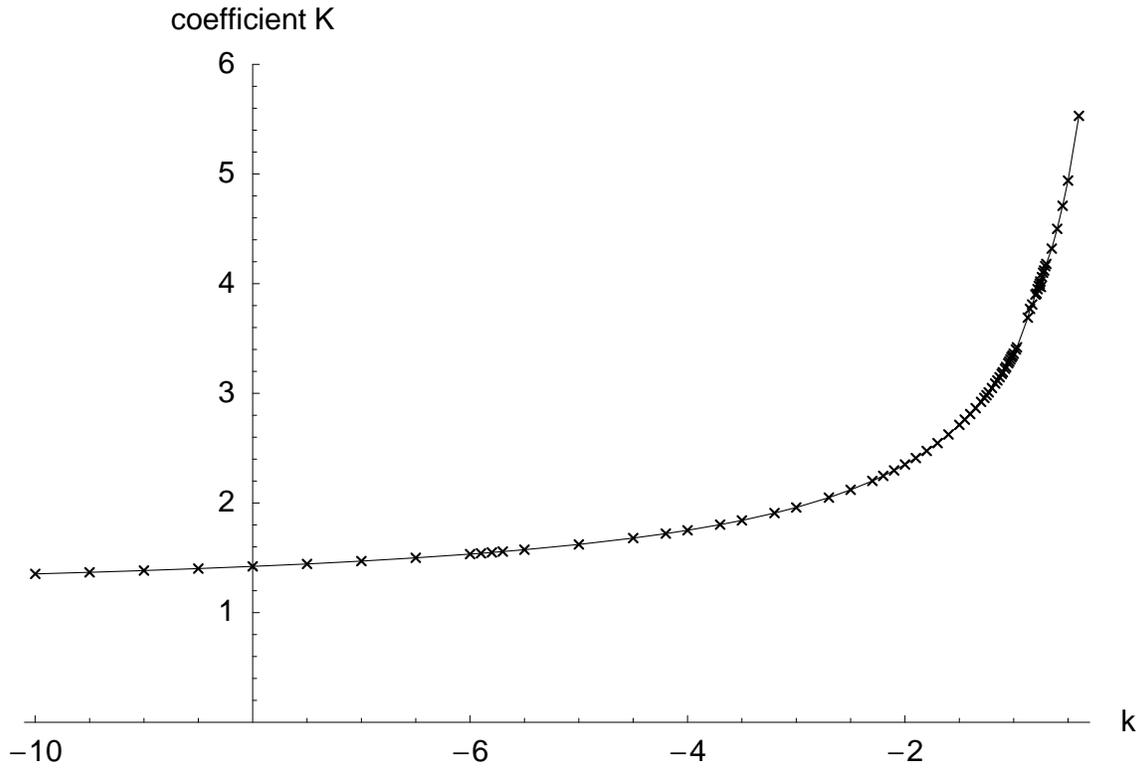}
\caption{Coefficient $K$ for the final Lorentz factor as defined in
  Eq.~\ref{final_gamma_propto} plotted as a function of density
  profile $k$. The crosses were computed from the cooling solution;
  the line connecting them is included to guide the eye.}
\label{finalgammacoeffs}
\end{figure}

We can find $\gamma_\mathrm{final}$ for a given fluid element directly
from our pre- and post-breakout solutions. To track the acceleration
of the fluid element while it is hot, we take the advective time derivative
of $\gamma$ in the pre- and post-breakout solutions for the hot fluid
and integrate with the proper limits. 
\begin{equation}
\frac{D\gamma}{Dt} = \frac{\gamma}{t}\frac{\left(\sqrt{3}-3\right)k}{g\chi-4-2\sqrt{3}}
\end{equation}
\begin{equation}
\frac{Dg\chi}{Dt} = \frac{1}{t}\frac{\left(2-g\chi\right)(g\chi-\gchinought)}{g\chi-4-2\sqrt{3}}
\end{equation}
\begin{equation}
\frac{D\gamma}{Dg\chi}=\gamma\,\frac{\left(\sqrt{3}-3\right)k}{(2-g\chi)(g\chi-\gchinought)}
\end{equation}
Before breakout, the fluid element's $g\chi$ goes from $g\chi=1$ when it
is shocked to $g\chi\rightarrow -\infty$ at breakout. After breakout, 
the fluid element's $g\chi$ goes from $g\chi\rightarrow\infty$ to
$g\chi\simeq \gchicold$. So when the fluid becomes cold, we have
\begin{equation}
\gamma=\gamma_\mathrm{shocked}(\gchinought-1)^{\frac{\left(\sqrt{3}-3\right)k}{\gchinought-2}}\left(\frac{g\chi-2}{g\chi-\gchinought}\right)^{\frac{\left(\sqrt{3}-3\right)k}{\gchinought-2}} \;\;\; .
\end{equation}
To find the correct $g\chi$ at which to evaluate the above, we set
$C=\gamma f/h$ to be the temperature (up to a factor of
3) at the $g\chi$ of interest and use Eqs.~\ref{f_hot}, \ref{h_hot}
to express $(g\chi-2)/(g\chi-\gchinought)$ in terms of $C$. This gives
\begin{equation}
\gamma=C^{-\sqrt{3}}\gamma_\mathrm{shocked}^{\sqrt{3}+1} \;\;\; .
\end{equation}
To this we add the extra factor given by the cooling solution to get
the coefficients $K$ shown in Figure~\ref{finalgammacoeffs}. In
particular,
\begin{eqnarray}
\gamma_\mathrm{final}=1.96\gamma_\mathrm{shocked}^{\sqrt{3}+1}&
\;\;\;\;\;\;\;\;\;& k=-3 \\
\gamma_\mathrm{final}=2.71\gamma_\mathrm{shocked}^{\sqrt{3}+1}&
\;\;\;\;\;\;\;\;\;& k=-3/2 \;\;\; .
\end{eqnarray}
This result is close to the results of \cite{tan01}, who find a
coefficient of $\sim 2.6$ when $k=-3$. Note that $p/n=1$ corresponds
to $C^{-\sqrt{3}}=0.149$: $\gamma$ grows by a factor of $\sim 15$
after the fluid element becomes nominally cold.

\begin{figure}
\plotone{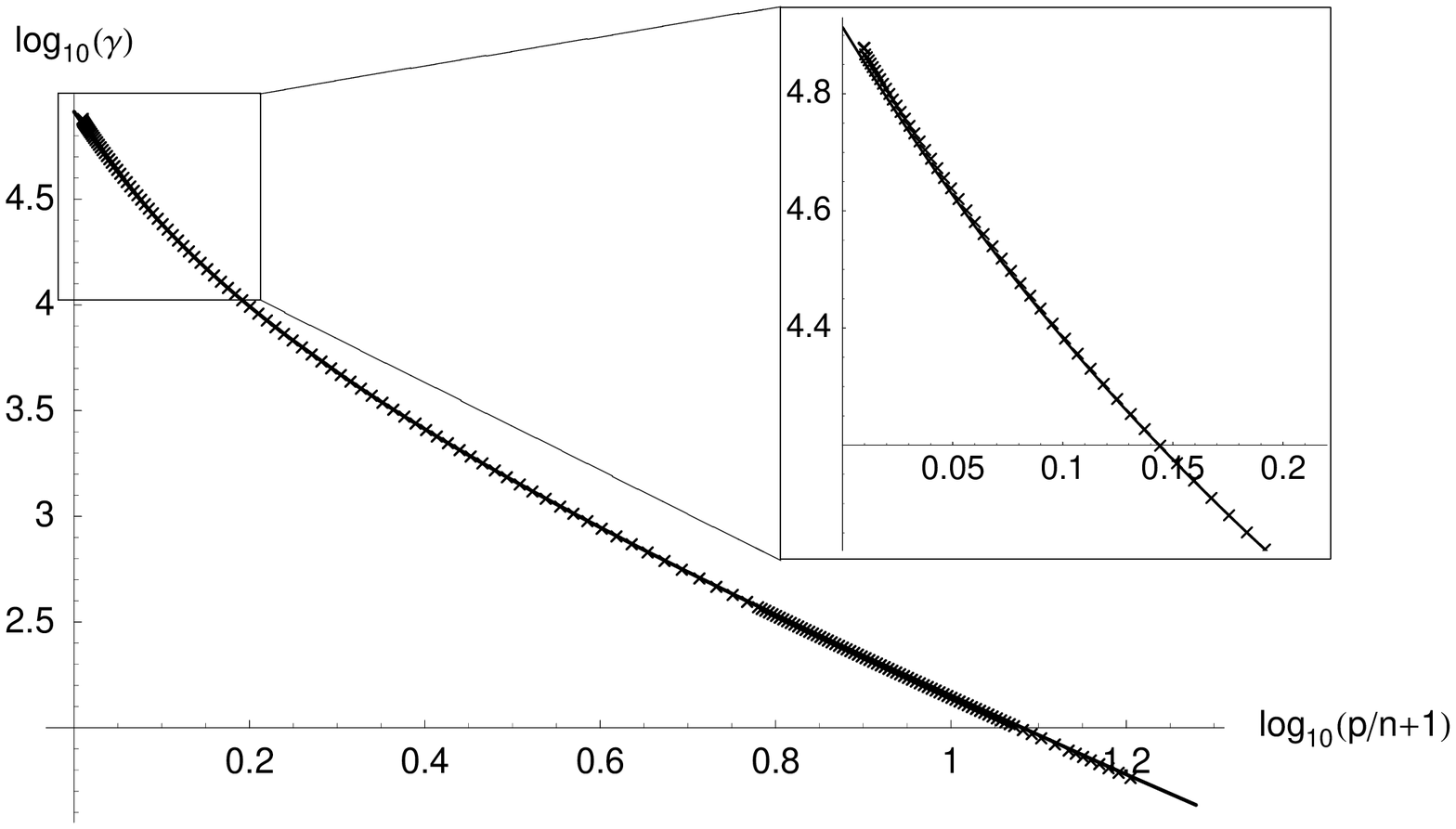}
\caption{Lorentz factor $\gamma$ of a single fluid element as a
  function of the temperature $p/n$ of that fluid element. Lower
  temperatures and later times are towards the left. The solid line is
  the self-similar solution; the points are the results of numerical
  simulations. Both calculations were done for $k=-3$. As the fluid
  element becomes cold, the evolution of its $\gamma$ with $p/n$
  deviates from the power law seen at high temperatures.}
\label{gpn_profiles}
\end{figure}

The growth of $\gamma$ as a function of the temperature for a single
fluid element is shown in Figure~\ref{gpn_profiles}, which also shows
good agreement between the cooling solution and direct numerical
simulations of the hydrodynamic equations. Because the Lorentz factors
near the front of the flow in particular become very large at late
times, it is difficult to produce numerical simulations that remain
accurate as the fluid cools all the way to $p/n\ll 1$. As a result,
the numerical simulation shown in Figure~\ref{gpn_profiles} cuts off
while the fluid Lorentz factor is 9\% smaller than the final Lorentz
factor predicted by the cooling solution.

\section{Effects of spherical geometry} \label{spherical}

We can estimate the ranges of initial positions and Lorentz factors
for which corrections to our planar solutions due to the star's
spherical geometry are important. Spherical geometry significantly
affects a given fluid element's acceleration if the distance between
the fluid element and the star's center doubles before the fluid
element finishes accelerating. Once the fluid element has traveled a
distance comparable to $R_*$, the star's radius, it has expanded
significantly in directions perpendicular to its motion; planar
solutions cannot account for this transverse expansion.  For our
estimate we therefore check which fluid elements have cooled to
$p/n\sim 1$ by time $t\sim R_*$.

Consider a fluid element whose position before being shocked is
$x_\mathrm{init}<0$. According to the relativistic shock jump
conditions, immediately after this fluid element is shocked it has
\begin{equation}
\left(\frac{p}{n}\right)_\mathrm{init}
 \sim \gamma_\mathrm{init}
 \sim \Gamma
 \sim \left(\frac{x_\mathrm{init}}{x_\mathrm{rel}}\right)^{-m/2}
\end{equation}
where $\Gamma$ is the shock Lorentz factor in the pre-breakout
solution and $x_\mathrm{rel}$ is the position of the shock when it
first becomes relativistic, or when $\Gamma\simeq \sqrt{2}$. Right
after breakout our fluid element follows the hot solution; when it
reaches $\chi=1$, it has accelerated and expanded by a factor of order
unity since being shocked, so it still has
$p/n\sim(x_\mathrm{init}/x_\mathrm{rel})^{-m/2}$. Also, at this point
$\Gamma\sim\gamma_\mathrm{init}$ so the time is$
\sim|x_\mathrm{init}|$. We track the temperature of the fluid element
over time until it reaches $p/n=1$ by taking the advective time
derivative of $p/n$:
\begin{equation}
\frac{D\ln (p/n)}{D\ln t}
 = \frac{k\left(\sqrt{3}-1\right)}{g\chi-4-2\sqrt{3}}
 \simeq \frac{\sqrt{3}-3}{6}
\end{equation}
where the final equality holds while the fluid element is at
$g\chi-\gchinought\ll\gchinought$ ---that is, near the front of the
flow, where most of the acceleration takes place. We then have
\begin{equation}
\left(\frac{p}{n}\right)
 \sim \left(\frac{p}{n}\right)_\mathrm{init}
      \left(\frac{t}{|x_\mathrm{init}|}\right)^{(\sqrt{3}-3)/6} \;\;\; .
\end{equation}
We now impose the condition $p/n\sim 1$ at or before $t\sim R_*$ to get
the scaling
\begin{equation}
\frac{|x_\mathrm{rel}|}{R_*} 
 < \left(\frac{|x_\mathrm{init}|}{R_*}\right)
         ^{1+\frac{1+\sqrt{3}}{3k}} \;\;\; .
\end{equation}
Sphericity corrections are unimportant for fluid elements which
satisfy this condition.  For $0>k\geq -\left(1+\sqrt{3}\right)/3$, the
exponent on the right hand side is not positive, so spherical
corrections are unimportant as long as $|x_\mathrm{rel}|<R_*$.  On the
other hand, for $k<-\left(1+\sqrt{3}\right)/3$ spherical corrections
are important for a layer of fluid initially adjacent to the star's
surface: in our self-similar solutions for these $k$ values, fluid
elements that start arbitrarily close to the star's surface will take
arbitrarily long to cool.

The equivalent condition on $\gamma_\mathrm{init}$, the Lorentz factor
of a given fluid element just after being shocked, is
\begin{equation}
\frac{|x_\mathrm{rel}|}{R_*} 
 < \left(\gamma_\mathrm{init}\right)
    ^{-\frac{1+\sqrt{3}+3k}{\left(3+\sqrt{3}\right)k}} \;\;\; ;
\end{equation}
the one for $\gamma_\mathrm{final}$, the fluid element's Lorentz factor
after it finishes accelerating, is
\begin{equation}
\frac{|x_\mathrm{rel}|}{R_*} 
 < \left(\gamma_\mathrm{final}\right)
    ^{-\frac{1+\sqrt{3}+3k}{\sqrt{3}k}} \;\;\; .
\end{equation}

\section{Summary and discussion} \label{summary}

We have derived a new self-similar solution, the cooling solution, for
the flow that results when a relativistic shock breaks out of a
polytropic envelope. The cooling solution is based on our
identification of the characteristic position with the point where the
fluid cools to nonrelativistic temperatures. The cooling solution
shows that the transition between hot and cold fluid in the flow is
self-similar even though this transition is not included---indeed, is
not self-similar---in the old post-breakout solution for the hot fluid
alone. We present a description of a non-self-similar flow using a
composite of two distinct self-similar solutions in which the time
evolution and physical interpretation of the characteristic scales
differ.  As the envelope's original density profile $k$ increases past
$k=-\left(1+\sqrt{3}\right)/3$, the flow dictated by the composite
solution changes qualitatively from one where the slowest fluid at the
back of the flow cools most quickly to one where the fastest fluid at the
front of the flow cools most quickly.

The cooling solution allows accurate calculation of the final Lorentz
factors of the shocked fluid elements. Given a stellar model for a
core-collapse supernova progenitor and an input explosion energy, we
can extract the initial density profile of the progenitor envelope and
the initial shock velocity and use the cooling solution to find the
Lorentz factor and kinetic energy profiles in the relativistic ejecta
after the ejecta finish accelerating. This provides an accurate value
for the energy available to produce observable lightcurves via
interaction between the ejecta from the model explosion and the
progenitor's surroundings.

\cite{kikuchi07} also investigate cooling in the flow produced after a
relativistic shock breakout. They focus on a $k=-3$ density profile
and use as starting point the work of \cite{nakayama05}, who found a
numerical self-similar solution for the hot planar flow in a
Lagrangian framework. While \cite{kikuchi07} also describe the cooling
flow with a system of ODEs, they conclude that no self-similar
solution exists for the cooling flow: instead of adopting a new
characteristic scale in writing the ODEs for the cooling flow, they
retain the characteristic scale relevant to the hot solution. Because
they do not recognize the self-similarity of the flow, and because
they do not integrate until the fluid has cooled enough, they find
that initially hot fluid does not stop accelerating. As a result they
cannot express the final Lorentz factors of given fluid elements in
terms of the initial ones.

In addition, \cite{kikuchi07} focus on corrections to their planar
flow due to spherical geometry. They perturb the hydrodynamical
equations to lowest order in a new variable equal to the position
coordinate scaled to the stellar radius. Their non-self-similar
solutions to the perturbed equations agree well with their numerical
simulations only at very early times, before the fluid has expanded by
a factor of $\sim$2. They look at sphericity effects in the cooling
flow using numerical simulations and find that for fluid whose initial
fractional distances from the star's surface to its center lie between
about $4\times 10^{-4}$ and $3\times 10^{-3}$, sphericity effects
change the final velocities by factors of order unity in simulations
with $k=-3$, $\Gamma=10^{5}$ and $10^{6}$ at $t=1$, and $R_*=3.3$ in
units where $c=1$. Although such a fast shock is unrealistic since it
would have $\Gamma\sim 10^4$ or $10^5$ at the star's center, our
scaling in \S\ref{spherical} also indicates that sphericity
corrections would be important for these parameters.

\cite{tan01} also consider sphericity corrections for the density
profile $k=-3$, but they treat the opposite limit of mildly
relativistic shocks with $\Gamma-1$ ranging from about $7\times
10^{-3}$ to $0.28$ when the shock is halfway from the star's center to
its surface.  Only their two most energetic numerical calculations of
sphericity effects attain shock Lorentz factors $\Gamma>\sqrt{2}$, so
nonrelativistic estimates of sphericity corrections are relevant for
most of the regime they consider. However, those two most energetic
shock calculations show that for the relativistic fluid elements
shocked to initial Lorentz factors between about $\sqrt{2}$ and $4$,
sphericity decreases the final velocities by twenty to thirty
percent. Our scaling in \S\ref{spherical} indicates that for those two
scenarios, sphericity corrections should be important for fluid
elements shocked to Lorentz factors larger than about 1.7 and 2.4
respectively; this is roughly consistent with the findings of
\cite{tan01}.

\acknowledgements{This research was partially funded by an IRG grant
and a NASA ATP grant. MP thanks the Association of Members of the
Institute for Advanced Study for support. RS is a Packard Fellow and
an Alfred P.~Sloan Research Fellow.}

\bibliographystyle{icarus}
\bibliography{shocks}
\nocite{pan06b}

\end{document}